\documentclass[twocolumn,aps,prb,showpacs]{revtex4}
\usepackage[latin9]{inputenc}
\usepackage{color}
\usepackage{amsmath}
\usepackage{amssymb}
\usepackage{graphicx}
\usepackage{esint}
\usepackage[unicode=true,pdfusetitle,
 bookmarks=false,
 breaklinks=false,pdfborder={0 0 1},backref=section,colorlinks=false]
 {hyperref}
\hypersetup{
 linkcolor=blue,urlcolor=blue,colorlinks,citecolor=blue}
\usepackage{breakurl}

\makeatletter
\@ifundefined{textcolor}{}
{%
 \definecolor{BLACK}{gray}{0}
 \definecolor{WHITE}{gray}{1}
 \definecolor{RED}{rgb}{1,0,0}
 \definecolor{GREEN}{rgb}{0,1,0}
 \definecolor{BLUE}{rgb}{0,0,1}
 \definecolor{CYAN}{cmyk}{1,0,0,0}
 \definecolor{MAGENTA}{cmyk}{0,1,0,0}
 \definecolor{YELLOW}{cmyk}{0,0,1,0}
}

\usepackage{epstopdf}\usepackage{dcolumn}% needed for some tables
\usepackage{bm}% for math
\usepackage{array}

\makeatother

\begin{document}
% the following line is for submission, including submission to the arXiv!!
%\hspace{5.2in} \mbox{Fermilab-Pub-04/xxx-E}

\title{Spontaneous Layer Polarization and Conducting Domain Walls in the 
Quantum Hall Regime of Bilayer Graphene}

\author{Kusum Dhochak$^1$, Efrat Shimshoni$^2$, and Erez Berg$^1$}

\affiliation{$^1$Department of Condensed Matter Physics$,$ Weizmann Institute of
Science$,$ Rehovot$,$ 76100$,$ Israel}

\affiliation{$^2$Department of Physics$,$ Bar Ilan University$,$ Ramat Gan$,$ 52900$,$
Israel}

\date{\today}
\begin{abstract}
Bilayer graphene subjected to perpendicular magnetic and electric
fields displays a subtle competition between different symmetry broken phases, resulting from an interplay between the internal
spin and valley degrees of freedom. The transition between different
phases is often identified by an enhancement
of the conductance. Here, we propose that the enhanced conductance at the 
transition is due to the appearance of robust conducting edge states at domain 
walls between the two phases. 
We 
formulate a criterion for the existence of
such conducting edge states at the domain walls. 
For example, for a spontaneously layer polarized state at
filling factor $\nu=2$, domains walls between regions of opposite
polarization carry conducting edge modes. A microscopic analysis shows
that lattice-scale interactions can favor such a layer polarized
state.
\end{abstract}


\maketitle

\begin{figure}
\includegraphics[width=1\columnwidth]{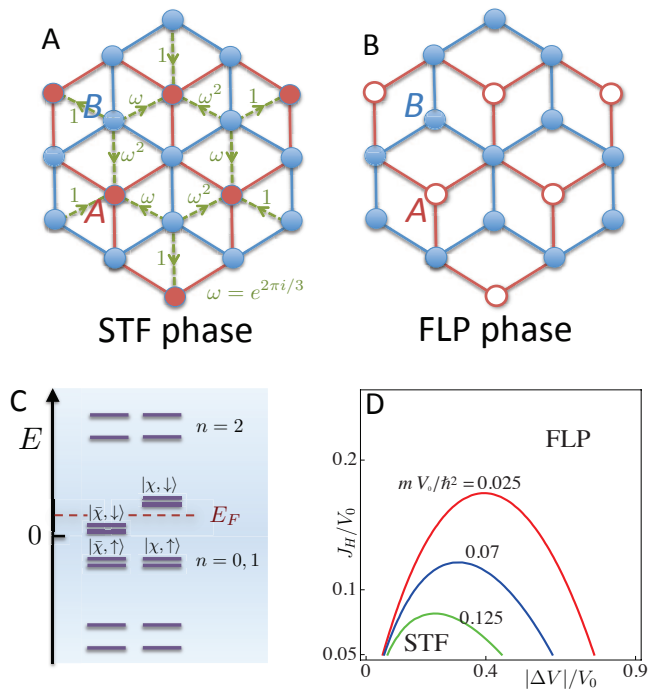}\caption{(A) The staggered flux 
(STF) phase. In this phase, translational symmetry is spontaneously broken, and 
there is a spontaneous staggered flux between the two graphene sheets. The order 
parameter can be described as a complex hopping amplitude between the A sublattice of the bottom layer and the B sublattice of the top layer. (B) The 
fully layer polarized (FLP) phase, which breaks inversion symmetry 
spontaneously. The electrons in the $n=0,1$ Landau levels with spin 
antiparallel to the external magnetic field occupy a single valley (and a single 
layer). (C) Energies of the Landau levels
at filling factor $\nu=2$. Ignoring the Zeeman splitting and electron-electron
interactions, there are eight degenerate zero-energy Landau levels,
corresponding to spin, valley, and orbital ($n=0,1$) indices. The exchange 
interactions favour a ferromagnet in spin and valley manifold with equal 
occupancy of $n=0,1$ orbitals. The
Zeeman coupling picks a direction for spin and splits the spin degeneracy. 
Finally, exchange interactions spontaneously 
split the remaining valley degeneracy, favoring a spinor 
$\vert\bar{\chi}\rangle$
in valley space over the orthogonal spinor $\vert\chi\rangle$. (D) Ground 
state mean-field phase diagram for $\nu=2$ as a function of $J_H$ and $\Delta V$ 
(see Eq.~\ref{eq:Hex}). The other coupling constant 
were fixed as follows: $U=V_1=V_0$ and $V_2=V_3$. The phase boundaries between the FLP 
and the STF phases are shown for different dimensionless coupling strengths: 
$mV_0/\hbar^2=0.025, 0.07,  0.125$. \label{fig1}}
\end{figure}

\section{Introduction}
Bilayer graphene (BLG) is a rich playground
to explore many-body physics\cite{Kotov2012}. At zero magnetic field, the energy bands
exhibits a quadratic touching that can lead to a variety of many-body
instabilities\cite{Sun2009,Zhang2010, Vafek2010,Vafek2010a, 
Nandkishore2010,Lemonik2010,zhang2011a,Lemonik2012,Zhang2012}. Signatures of a 
symmetry-broken
ground state at $B=0$ have been observed experimentally\cite{Martin2010, Weitz2010,Mayorov2011,Velasco2012,Freitag2012},
although the precise nature of this state is still debated. When a
magnetic field is applied perpendicular to the system, the Landau
levels are highly degenerate, including spin, valley, and (for the
zeroth Landau level) also an orbital degree of freedom \cite{McCann2006}.
This degeneracy can be lifted by exchange interactions\cite{Feldman2009, Zhao2010, Weitz2010,Freitag2012,Velasco2014}, leading to
different kinds of broken symmetry states \cite{Jung2011a, Kharitonov2012, 
Kharitonov2012a}. Which state is favored depends on the
nature of the lattice-scale interactions between electrons, which
break the approximate SU(4) symmetry in spin and valley space.

In BLG, an electric field perpendicular to the plane couples to the valley
degree of freedom of the Landau levels. Upon tuning the strength
of the magnetic and electric perpendicular fields ($B$ and $E$,
respectively) at a fixed filling fraction, transitions between different
ordered states can be induced\cite{Weitz2010,Velasco2014,luo2014a,luo2014}. 
These transitions are identified
by peaks in the conductance along lines in the $(E,B)$ plane. The
mechanism for this enhanced conductance at the transitions remains unexplained. These transitions are expected to be of first order in the clean limit; they are
described as a level crossing of different ground states, without
closing of the energy gap above these two states, and hence there is no obvious reason for an enhancement of the conductance.

In this paper, we propose that the enhanced conductance at the transitions
can result from the appearance of robust one-dimensional conducting
modes at domain walls between different phases. The possibility of
the appearance of such (non-chiral) modes, either at the edge of the sample or
at domain walls, has been proposed in Refs. 
\cite{Jungwirth2001,Fertig2006,Martin2008,Abanin2006,Killi2010,Mazo2011,
zhang2011a,zhang2011b, Mazo2012,Vaezi2013a,Zhang2013,Li2014,Abanin2010}. 
These edge states are partially protected against backscattering
by the approximate conservation of either the spin or pseudospin (valley)
quantum numbers, and are robust in the presence of electron-electron
interactions. Evidence for such edge state have been observed under a high 
in-plane magnetic field\cite{Maher2013}. We formulate a simple criterion for 
robust edge states at domain walls between
two quantum Hall ferromagnetic phases with the same filling fraction,
based on their symmetry properties and their quantum numbers.

As an example, we analyze the case of $\nu=2$ at $E=0$. In this
case, to leading approximation, the partially filled Landau levels
have an SU(2) valley (pseudospin) degree of freedom \cite{Barlas2008,Abanin2009}.
This symmetry is broken either by an applied electric field or by
lattice scale interactions. We argue that the experimental findings
of Wietz \textit{et. al.}~\cite{Weitz2010} are consistent with a spontaneously 
layer polarized
phase {[}an easy axis ferromagnet, in terms of the SU(2) pseudospin{]}.
The domain walls between regions of opposite polarization support
conducting edge modes\cite{Fertig2006,Mazo2011}. At $E=0$, the domain walls
percolate, leading to an enhanced conductance. We present a Hartree-Fock
analysis of a microscopic model, and demonstrate how such a layer
polarized phase can be favored over other possible broken-symmetry states in the physically relevant regime of parameters. 
Interestingly, to obtain such a phase,
it is essential to treat all the Landau levels explicitly, rather
than projecting to the partially filled zero-energy Landau levels. 

\section{Setup}
We consider a BLG sheet in the Bernal stacking.
The low-energy single particle effective Hamiltonian is
written as~\cite{McCann2006}
\begin{equation}
H_{0}=-\frac{1}{2m}\left(\begin{array}{cc}
0 & \left(\pi_{x}-i\tau^{z}\pi_{y}\right)^{2}\\
\left(\pi_{x}+i\tau^{z}\pi_{y}\right)^{2} & 0
\end{array}\right).\label{eq:H0}
\end{equation}
Here, $m$ is the effective mass of the bands near zero energy, and
$\pi_{i}=-i\partial_{i}-eA_{i}$ ($i=x,y$) where $\vec{A}$ is the
vector potential. $\vec{\tau}$ are Pauli matrices acting on the valley
index, such that $\tau^{z}=\pm1$ corresponds to the $\pm K$ point
in momentum space, where $K=(-4\pi/3\sqrt{3}a_0,0)$, and $a_0$ is the
inter-atomic spacing within each layer. We set the units such that
$\hbar=c=1$. The $2\times2$ Hamiltonian acts on the spinor $(\psi_{A},\psi_{B})$,
where $\psi_{A(B)}$ annihilates an electron on sublattice $A$ ($B$)
in the bottom (top) layer, respectively. We define the 8-component
spinor $\Psi$, that contains annihilation operators in layer $\mu^{z}=\pm1$,
valley $\tau^{z}=\pm1$, and spin $s^{z}=\pm1$.

In the presence of a uniform orbital magnetic field, the single particle spectrum
consists of a series of Landau levels whose energies are $E_{n}=\pm \omega_c \sqrt{n(n-1)}$, where $\omega_c \equiv eB/m$ and
$n=0,1,\dots$~\cite{McCann2006}. 
Each Landau level is four-fold
degenerate, with two possible valley labels, $\tau^{z}=\pm1$, and
two spin labels, $s^{z}=\pm1$ (neglecting the Zeeman splitting).
In addition, the $n=0,1$ Landau levels are degenerate.

We write the full Hamiltonian as

\begin{equation}
\hat{H}=\hat{H}_{0}+\hat{H}_{Z}+\hat{H}_{C}.\label{eq:H}
\end{equation}
Here, $\hat{H}_{0}=\int d^{2}r\Psi^{\dagger}(r)H_{0}\Psi(r)$, $\hat{H}_{Z}=-g\mu_{B}B\int d^{2}r\Psi^{\dagger}s^{z}\Psi$
is the Zeeman coupling, and $\hat{H}_{C}$ is the Coulomb interaction
(to be discussed below).

When some of the zero-energy Landau levels are empty, the 
system tends to form a quantum Hall ferromagnetic state which breaks the 
symmetry in spin
and valley space, in order to gain Coulomb exchange energy. At the
lowest Landau level, where there is an additional orbital ($n=0,1$)
degeneracy, maximum exchange is gained by filling the $n=0,1$ orbitals
together with the same state in $\tau$, $s$ space~\cite{Barlas2008,Abanin2009}.
We will assume that this form of ``Hund's rule'' is obeyed below,
although it is not essential for the general criterion for conducting
edge states between different phases.

The filling fraction $\nu$ is defined as the number of electrons
per flux quantum, with respect to the charge neutral state. The $\nu=0$
quantum Hall state is determined by two orthogonal spinors in spin/valley
space, $\chi_{1}$ and $\chi_{2}$, such that of the eight degenerate
zero-energy Landau levels, four are occupied: $\vert\chi_{i},n\rangle$
with $i=1,2$ and $n=0,1$~\cite{Kharitonov2012}. Similarly, the $\nu=2$
state is determined by a single spin/valley spinor $\chi$, such that
of the $E=0$ Landau level states, only the states $\vert\chi,n\rangle$
($n=0,1$) are \emph{empty} (see Fig.~\ref{fig1}C). At a given filling fraction, there is
a manifold of possible states; this degeneracy is lifted by the Zeeman
field, and applied electric field perpendicular to the sheet (which
breaks the degeneracy between the layers), and by the short-range
exchange interactions.

\section{Condition for conducting modes at domain walls}
As external
parameters are varied at a fixed filling fraction, the system can
undergo phase transitions between different ordered states in spin/valley
space. If disorder effects are ignored, these transitions are generically of
first order. At the transition point, we expect a phase mixture of
two phases.

Consider a domain wall between two such phases. Here, we discuss a
sufficient condition for the appearance of conducting edge states
at the domain wall. Our condition is formulated as follows: suppose
that the two phases on either side of the domain wall are invariant
under a common $U(1)$ symmetry generated by an operator $\hat{G}$ in spin and/or
valley space. (The symmetry could be generated by $s^{z}$, $\tau^{z}$,
or some combination of the two.)\emph{ }Define the weighted filling
fraction \emph{$\tilde{\nu}=\sum_{j\in\mathrm{filled}}q_{j}$,} where
$j$ runs over the filled states, and $q_{j}$ is the charge of the
state $j$ under $\hat{G}$. \emph{If $\tilde{\nu}$ of the two phases is
different, there is necessarily a gapless edge state at the domain
wall between them.} This edge state is robust in the presence of arbitrary interactions
and disorder, as long as they preserve $\hat{G}$. For instance, if $\hat{G}=\tau^{z}$,
the edge states will be protected as long as we neglect lattice-scale
disorder that causes inter-valley scattering, and the domain wall
itself is sufficiently smooth. The conductance of the edge state is
given by $\Delta\tilde{\nu}e^{2}/2h$, where $\Delta\tilde{\nu}$
is the difference of $\tilde{\nu}$ between the two phases.

A simple way to understand the existence of a gapless edge state is
to define the Hall conductance related to the conserved $U(1)$ charge
under $\hat{G}$. We introduce a vector potential $\vec{A}_{G}$ that couples
to the charge under $\hat{G}$, by substituting $-i\vec{\partial}\rightarrow-i\vec{\partial}-\hat{G}\vec{A}_{G}$
in $\hat{H}_{0}$. The response of the current $\vec{j}_{G}=\partial\hat{H}/\partial\vec{A}_{G}$
to an applied electric field in the plane, described by the ``$\hat{G}$
Hall conductance'' $\sigma_{G}^{H}$, is quantized (this is 
analogous to the spin Hall conductance in a quantum spin Hall state
with conserved spin~\cite{Kane2005, Bernevig2006}). Phases with different 
$\tilde{\nu}$
correspond to different $\sigma_{G}^{H}$, and must have gapless edge
states between them. This argument is expected to hold in the presence
of arbitrary interactions and disorder, as long as the symmetry $\hat{G}$
is preserved and the gap in the bulk of both phases is maintained.

For example, consider the state with filling factor $\nu=2$, specified by
the four-component spinor $\chi$ defined above. The Zeeman coupling
favors a particular spin component, say $s^{z}=1$. To specify the state,
the spinor $\chi$ in valley space remains to be determined. Lattice scale
exchange interactions will select either a spontaneously layer-polarized
state with $\langle \vec{\tau} \rangle \parallel \hat{z}$, or an in-plane polarized phase with $\langle \vec{\tau} \rangle \perp \hat{z}$. 
If
the layer-polarized state is favored, then domain walls between the
$\tau^{z}=\pm1$ phases carry non-chiral edge states. These edge states are 
robust as long as we can neglect inter-valley scattering (which requires the 
domain wall to be smooth on the lattice scale; see next section for a discussion of the characteristic length scale of the domain walls), and remain so for arbitrary 
interactions. The edge states have conductance of $2e^{2}/h$ (where the factor 
of $2$ is due to the orbital degeneracy). The in-plane state
breaks a continuous symmetry; in this phase, there are no sharp
domain walls. If weak inter-valley scattering disorder is present,
the valley polarization is locally pinned by the disorder, and twists
gradually in space.

Physically, the in-plane polarized state breaks translational symmetry spontaneously, while the valley polarized state breaks inversion symmetry. On the lattice scale, the former is described as a ``staggered flux" (STF) state, and the latter is a ``fully layer polarized'' (FLP) state (see Fig.~\ref{fig1}A,B).

\section{Hartree-Fock analysis for $\nu=2$}
 The long-range part of the 
Coulomb interactions is symmetric in spin
and valley space, and therefore it does not lift the degeneracy between
the layer-polarized and the in-plane polarized states. The degeneracy
is lifted by short-range (lattice scale) exchange interactions. In
order to analyze the competition between these phases, we use the
following form for the short-range exchange Hamiltonian:\begin{widetext}

\begin{align}
\hat{H}_{\mathrm{ex}} & =\int d^{2}r\left\{ 
V_{0}n^{2}+\sum_{\mu}\left[\sum_{\tau}Un_{\mu\tau\uparrow}n_{\mu\tau\downarrow}
+V_{1}n_{\mu,K}n_{\mu,K'}-J_{H}\vec{S}_{\mu,K}\cdot\vec{S}_{\mu,K'}\right]+\sum_
{\tau}V_{2}n_{1,\tau}n_{-1,\tau}+V_{3}\sum_{\mu}n_{\mu,K}n_{-\mu,K'}\right\} 
.\label{eq:Hex}
\end{align}

\end{widetext}Here, $n_{\mu\tau s}=\psi_{\mu\tau s}^{\dagger}\psi_{\mu\tau 
s}^{\vphantom{\dagger}}$ 
($\mu=\pm1$, $\tau=K,K'$, $s=\uparrow,\downarrow$ are layer, valley,
and spin indices, respectively, and we have suppressed the spatial
argument $\vec{r}$ for brevity), $n_{\mu\tau}=\sum_{s}n_{\mu\tau s}$,
$n=\sum_{\tau,\mu}n_{\tau\mu}$, and $[\vec{S}_{\mu\tau}]_{s,s'}=\psi_{\mu\tau s}^{\dagger}\vec{\sigma}_{ss'}\psi_{\mu\tau s'}^{\vphantom{\dagger}}$
where $\vec{\sigma}$ are Pauli matrices. $V_{0}$ is the part of
the Coulomb interactions which are isotropic in layer, valley, and
spin space; $U$ is the strength of the interaction between two electrons
in the same layer and valley; $V_{1}$ is an inter-valley, same-layer
coupling constant; $J_{H}$ is an inter-valley Hund's rule coupling
constant; and the $V_{2}$ ($V_{3}$) terms describe inter-layer,
intra- (inter-) valley interactions, respectively. For simplicity,
we have assumed that the spatial overlap between wavefunctions of
electrons in different layers is small, so we can neglect inter-layer
exchange interactions. On general grounds, we expect the following
relations between the interactions: $V_{0}\gtrsim U\ge V_1>J_{H}>0$,
$U>V_{2}\ge V_{3}>0$. (For more discussion of the microscopics of the exchange 
interactions, see Appendix~\ref{apdx:BB}.)

We now proceed to perform a mean-field analysis on the Hamiltonian
$\hat{H}=\hat{H}_{0}+\hat{H}_{Z}+\hat{H}_{\mathrm{ex}}$. Interestingly,
upon projection to the partially filled Landau level, the exchange
Hamiltonian (\ref{eq:Hex}) does not lift the degeneracy between the
layer-polarized and the in-plane polarized states at $\nu=2$. This
is because the partially filled Landau level is fully polarized in
spin and valley space; therefore, $\langle\hat{H}_{\mathrm{ex}}\rangle=0$
independent of the direction of polarization. In the following, we
avoid projecting to the lowest Landau level, and treat the entire
Landau spectrum. The contribution of the $n>1$ Landau levels to the
susceptibility in the particle-hole channel is divergent in the limit
$B\rightarrow0$, due to the quadratic band touching in the underlying
dispersion. This implies that the contribution of the higher
(occupied and unoccupied) states to the energetics is significant.

To estimate the ground state energies of the different states, we
use a variational Hamiltonian of the form

\begin{equation}
\hat{H}_{\mathrm{MF}}=\hat{H}_{0}+\hat{H}_{Z}-\int 
d^{2}r\sum_{\alpha,\beta,\zeta}\lambda_{\alpha\beta}^{(\zeta)}\Psi^{\dagger}
\mu^ { \alpha}\tau^{\beta}\frac{1+\zeta s^{z}}{2}\Psi,
\end{equation}
where $\alpha,\beta=0,x,y,z$ and $\zeta=\pm1$. The parameters 
$\lambda_{\alpha\beta}^{(\zeta)}$
are chosen to minimize $\langle\hat{H}\rangle_{\mathrm{MF}}$ (see aApendixes 
\ref{apdx:C} and \ref{apdx:D} for details).

Symmetry can be used to reduce the number of variational parameters.
We assume that none of the candidate states break spin rotational symmetry
around the $z$ axis. The fully layer polarized (FLP) state breaks
lattice inversion symmetry, represented by $\hat{I}=\mu^{x}\tau^{x}\times(\vec{r}\rightarrow-\vec{r})$,
but preserves translational symmetry and three-fold rotational symmetry,
$\hat{R}_{2\pi/3}=\exp(2\pi i\mu^{z}\tau^{z}/3)\times(\vec{r}\rightarrow\mathcal{R}\vec{r})$,
where $\mathcal{R}$ is a $2\times2$ rotation matrix (Appendix \ref{apdx:A}). The 
only mean-
field terms that are consistent with these symmetries are: $\{\mu^{z},\tau^{z},\mu^{z}\tau^{z}\}_{\mathrm{FLP}}$.
The staggered flux (STF) phase breaks translational symmetry, but
preserves $\hat{R}_{2\pi/3}$ around a certain three-fold axis and $\hat{I}$.
The allowed mean-field terms in this phase are $\{\mu^{x}\tau^{x},\mu^{y}\tau^{y},\mu^{z}\tau^{z}\}_{\mathrm{STF}}$.
Interestingly, the set of mean-field terms in the FLP phase are mapped
onto those of the STF phase under a unitary transformation given by
$\hat{U}=\exp(i\frac{\pi}{4}\mu^{x}\tau^{y})$, which also leaves $\hat{H}_{0}$
invariant. Therefore, interactions that are invariant under $\hat{U}$ {[}e.g.
the $V_{0}$ term in (\ref{eq:Hex}), or long-range Coulomb interactions{]}
do not lift the degeneracy between the FLP and STF phases.

The explicit evaluation of $\langle\hat{H}\rangle_{\mathrm{MF}}$
and the minimization over $\lambda_{\alpha\beta}^{(\zeta)}$ is tedious
but straightforward, and will be deferred to the appendixes 
\ref{apdx:C},\ref{apdx:D}, and \ref{apdx:E}. We will
quote some of the result here. The instability towards either an FLP
or STF phase occurs for any non-zero strength of the interactions,
as a result of the flatness of the Landau levels. The difference in
ground state energy between the FLP and STF states can be found analytically
in the limit of weak interactions. In this limit, the energy difference
per unit area $\Delta E\equiv E_{\mathrm{STF}}-E_{\mathrm{FLP}}$
is given by

\begin{align}
\Delta E & 
=\frac{\chi_{0}}{\ell_{B}^{4}}\left[\left(\frac{V_{0}}{2}+\frac{U}{4}\right)J_{H
}-\left(\frac{V_{0}}{2}+\frac{2V_{3}-U}{4}\right)\Delta V\right.\nonumber \\
 & \left.+\frac{1}{8}[2J_{H}^{2}+(U-V_{3})^{2}+2(\Delta V)^{2}]\right], 
\label{eq:DE}
\end{align}
where $\Delta V=V_1-V_{2}$, $\ell_{B}=\sqrt{\hbar/eB}$, and $\chi_{0}=\frac{m}{\pi}\sum_{n=2}^{N}\frac{\omega_{c}}{E_{n}}$
 ($N=E_c/\omega_{c}$, where $E_c$ is a high energy cutoff
of the theory). The ground state is the FLP state when $\Delta E>0$,
which occurs when $J_{H}$ is larger than a critical value of the
order of $\Delta V$. 

Figure~\ref{fig1}D shows the phase diagram as a function of $J_H$ and $\Delta 
V$. We have fixed the ratios between all the other coupling constants, and 
present results for different values of the dimensionless interaction parameter 
$mV_0$. For weak interactions [where Eq. (\ref{eq:DE}) applies], the FLP phase 
is favored for large $J_H$, whereas the STF phase is the ground state for small 
$J_H$ and an intermediate range of $\Delta V$. As the interaction strength 
increases, the region of the FLP phase expands (a naive estimate of the 
realistic interaction strength gives $mV_0\sim 1$). These findings do not 
depend sensitively on the precise values of the other coupling constants.

In the above analysis, we have disregarded the long-range part of the Coulomb 
interaction, and treated only contact (exchange) interactions (whose range is of 
the order of the short distance cutoff, $a \equiv 1/\sqrt{mE_c}$). As 
already noted, the $1/r$ part of the Coulomb interaction is symmetric in spin 
and valley space, and does not distinguish the FLP and STF phases. The 
dipole-dipole term, which falls as $1/r^3$, favours the STF phase (since it 
opposes layer polarization). However, a simple estimate shows that the dipolar 
energy per unit area, $E_{\mathrm{d}}\sim e^2d^2/\ell_B^5$ (where $d$ is the 
inter-layer spacing), is suppressed by a factor of $\sim d^2/a\ell_B \ll 1$ 
compared to the exchange energy difference between the two phases, 
Eq.~(\ref{eq:DE}) (taking $e^2 a$ as the typical magnitude of the exchange 
couplings). Therefore, the long-range dipolar interaction is typically 
negligible. 

 Finally, we comment on the structure of the domain walls between two oppositely 
polarized regions in the FLP phase. The ``easy axis'' anisotropy energy is of 
the order of $e^2 a$ per unit area, whereas the stiffness of the valley 
pseudospin $\vec{\tau}$ is $\sim e^2/\ell_B$. Dimensional analysis gives that 
the domain wall has a characteristic thickness of $\ell_{\mathrm{DW}} \sim 
\ell_B \sqrt{\ell_B/a} \gg a$. Therefore, we expect that inter-valley scattering 
induced by the domain wall is small.

\section{Discussion and relation to experiments}
Experimentally, an 
enhanced conductance was found along a line in the $(E,B)$ plane for filling 
factor $\nu=0$~\cite{Weitz2010}, and at $E=0$ for filling factor
$\nu=2$~\cite{Weitz2010,Velasco2014}. We interpret this enhancement of the conductance as arising from gapless edge modes at domain walls between two phases at the transition point. If these phases satisfy the criterion described above, then they are topologically distinct as long as the common $U(1)$ symmetry $G$ is preserved; therefore, there is a sharp phase transition between the two phases, even in the presence of interactions and disorder. The transport near the transition is then described in terms of percolation of the domain walls between the two phases~\cite{Chalker1988,Lee1993,Shimshoni1998,Kramer2005}. 

For $\nu=2$, as we have shown here, the enhanced conductance is readily explained if the $E=0$ ground state is spontaneously layer polarized. Our microscopic analysis shows how such a state can arise from short-range exchange interactions. Such spontaneous layer polarization can be detected directly by capacitance measurements~\cite{Young2011}. 

At $\nu=0$, the existence of an enhanced conductance at the finite $E$ transition allows us to put constraints on the nature of the states on either side of the transition. We assume that the $E=0$ ground state is a canted antiferromagnet~\cite{Kharitonov2012}. Then, according to our criterion, a domain wall with a partially layer polarized state~\cite{Kharitonov2012}, in which a coherent superposition of the two valley states is occupied, does not carry a protected edge mode, because the two phases do not have a common $U(1)$ symmetry. An edge state with a \emph{fully layer polarized} state, however, does have a conducting edge state, since the two phases have a common valley symmetry, and valley Hall conductance jumps across the domain wall.

\section{Acknowledgements}
We thank D. Abanin, A. H. Fertig, B. I. 
Halperin, C. Kane, G. Murthy, M. S. Rudner, A.
Yacoby, and A. Young for useful discussions. This work was partially supported 
by the US-Israel Binational Science 
Foundation (BSF) grant 2012120 (E.S.), and by the Israel Science Foundation 
(ISF) grants 231/14 (E.S.) and 1291/12 (E.B.). E. B. was also supported by the 
German-Israeli Foundation (GIF), the Minerva foundation, and a Marie Curie CIG 
grant.

\appendix

\section{Local Interaction Hamiltonian}\label{apdx:BB}
To write the interaction Hamiltonian, we start from a tight binding model with
the lattice structure of bilayer graphene. We are interested
in the local part of the low-energy effective interactions, which are 
anisotropic in
valley and layer space. We will assume weak coupling (small $e^2/v$), for which
we can simply project the microscopic (Coulomb) interactions

% \selectlanguage{english}%
\begin{equation}
H_{int}=\sum_{s,s'}\int
d^3\vec{r} \,
d^3\vec{r'}\, 
V(\vec{r}-\vec{r'})\varphi_{s}^{\dagger}(\vec{r})\varphi_{s}(\vec{r})\varphi_{
s'}^{ \dagger} (\vec{r'})\varphi_{ s' } (\vec{r'}),
\end{equation}
onto the low-energy subspace. Here, $\varphi_s(\vec{r})$ annihilates an 
electron 
at position $\vec{r}$ with spin $s$, and
$V(\vec{r}-\vec{r}')=e^{2}/|\vec{r}-\vec{r}'|$. We choose a basis of states 
whose support in momentum space is in the regions
$\left|\vec{k}-\vec{K}\right|<\Lambda$ and 
$\left|\vec{k}-\vec{K}'\right|<\Lambda$, where
$\Lambda$ is a momentum cutoff. In real space, these wavefunctions
are localized within a region of size $a\sim2\pi/\Lambda$. We assume
$a$ to be of the order of a few lattice constants.

Let us denote the basis functions by $\Phi_{\mu\tau}(\vec{r}-\vec{R}),$ 
where $\vec{R}$ is the center of mass of this orbital, $\vec{r}$ is a continuous
space variable, and $\mu,\tau$ are the layer and valley indices. $\psi_{\mu 
\tau 
s}$ is an operator that annihilates an electron in orbital $\Phi_{\mu \tau} 
(\vec{r}-\vec{R})$ with spin $s$.

The low-energy part of the field operators is given by 
\begin{equation}
\varphi_{s}(r)=\sum_{R,\mu,\tau}\Phi_{\mu\tau}(\vec{r}-\vec{R})\psi_{\mu\tau 
s}(\vec{R}).
\end{equation}
The interaction Hamiltonian takes the form

\begin{align}
 &H_{int} =\!\!\!\!\sum_{s,s',1,2,3,4}\!\!\left[\int
drdr'V(r-r')\Phi_{1}^{*}(\vec{r}-\vec{R}_{1})\Phi_{2}(\vec{r}-\vec{R}_{2}
)\right.\nonumber \\
 & \!\!\!\!\times \left. \Phi_{
3}^{*}(\vec{r}-\vec{R}_{ 3})\Phi_{4}(\vec{r}-\vec{R}_{4
})\right]
\psi_{1s}^{\dagger}(\vec{R}_{1})\psi_{2s}(\vec{R}_{2})\psi_{3s'}^{\dagger}(\vec{
R } _ { 3 } ) \psi_ {4s'} (\vec{R}_{4} ).
\end{align}
Here, we have used the short hand notation $1$ for 
$\{\mu_{1},\tau_{1},R_{1}\}$, 
etc.
The object within the square bracket is the coupling constant of the
particular operator
$\psi_{1s}^{\dagger}(\vec{R}_{1})\psi_{2s}(\vec{R}_{2})\psi_{3s'}^{\dagger}(\vec
{
R } _ { 3 } ) \psi_ {4s'} (\vec{R}_{4}) $.\\
\\

Calculating the values of the microscopic coupling constants from first 
principles is very difficult, because these coupling constants are strongly 
renormalized with respect to their bare values~\cite{Kharitonov2011}. We will 
mostly treat them as phenomenological parameters. Below, we make some 
physically-motivated simplifying assumptions, in order to reduce the number of 
independent parameters.

\subsection{Simplifying assumptions and explicit form of $H_{int}$}

\begin{enumerate}
\item We assume that $\vec{R}_{1}=\dots=\vec{R}_{4}.$This assumption is 
justified if
the orbitals $\Phi(\vec{r}-\vec{R})$ are sufficiently localized around 
$\vec{R}$. 
\item We consider two types of terms: intra-layer and inter-layer. We assume
that the overlap between orbitals localized in the two layers is negligible;
therefore, we will not consider terms that hop a pair from one layer
to the other. Moreover, the inter-layer interaction terms are spin
independent. 
\end{enumerate}
Within these assumptions, we get the following form for $H_{\mathrm{int}}$:

\begin{equation}
H_{\mathrm{int}}=H_{\mathrm{intra}}+H_{\mathrm{inter}}+V_0 (\psi^{\dagger} 
\psi)^2 \label{eq:hc_int}
\end{equation}

\begin{align}
H_{\mathrm{intra}}=&\sum_{\mu}\left( U(n_{\mu K\uparrow}n_{\mu 
K\downarrow}+n_{\mu 
K'\uparrow}n_{\mu
K'\downarrow})\right. \nonumber\\
& \left. +V_1 n_{\mu K}n_{\mu K'}-J_{H}\vec{S}_{\mu 
K}\cdot\vec{S}_{\mu K'}\right)
\end{align}

\begin{equation}
H_{\mathrm{inter}}=V_{2}(n_{1, K}
n_{-1, K}+ n_{1,K'}n_{-1,K'})+ \sum_{\mu} V_{3} n_{\mu,K} n_{-\mu,K'}
\end{equation}
with six parameters, $V_0,U,V_1,J_{H},V_{2},V_{3}$. On general grounds,
we expect the following inequalities to hold:
\begin{align}
&V_0\gtrsim U\geq V_1>J_{H},\nonumber\\
&V_1>V_{2}\geq V_{3}.
\end{align}
In addition to these local interactions, there are also long-range
Coulomb (monopole-monopole and dipole-dipole) interactions. 

\section{Mean Field Theory with Local Interactions: general 
formulation}\label{apdx:C}
Consider the following Hamiltonian:
\begin{equation}
H=H_{0}+H_{\text{int}}-\mu(\psi^{\dagger}\psi-n_{0}),
\end{equation}
where $H_0$ is the single particle Hamiltonian of BLG, $H_\mathrm{int}$ is a 
local exchange interaction (Eq.~\ref{eq:hc_int}). The chemical potential $\mu$ 
is chosen such that the density is $\langle\psi^{\dagger}\psi\rangle=n_{0}$.

We use a variational mean-field Hamiltonian: 
\begin{equation}
H_{\mathrm{MF}}=H_{0}-\sum_{a=1}^{15}\lambda_{a, 
s}\psi_{s}^{\dagger}O_{a}\psi_{s}-\mu_{0s}\psi^{\dagger}_{s}\psi_{
s}. \label { eq:HMF }
\end{equation}

 $O_{a}$ are the following matrices in valley and layer space:
\begin{align}
O_{a=0,..,15}= &\nonumber\\
&\{1, \mu^z \tau^z,\mu^{x},\mu^{y}\tau^{z}, 
 \,\mu^{z},\mu^{y}\tau^{y},\mu^{y}\tau^{x},\tau^{z} 
,\nonumber\\
 &\hspace{3mm}\mu^{x}\tau^{x},\mu^{x}\tau^{y},\mu^{x}\tau^{z},\mu^{y},\tau^{x},
\tau^{y},\mu^{
z}\tau^{x},\mu^ {z}\tau^{y}\}.
\end{align}

These form a complete basis of hermitian 
matrices in the layer and valley space. They satisfy
\begin{equation}
\mbox{Tr}O_{a}O_{a'}=\delta_{a,a'}.
\end{equation}

The mean-field energy is
\begin{align}
& E(\{\lambda_{a}\},\mu_{0})=\langle H\rangle\nonumber\\
& \hspace{2mm}=\langle
H_{MF}\rangle+\sum_{a=1}^{15}\lambda_{a s}\langle \psi^{\dagger}_s
O_{a} \psi_s\rangle+\mu_{0s}\langle\psi^{\dagger}_s 
\psi_s\rangle\nonumber\\
& \hspace{2mm}
+\langle H_{\mathrm{int}}\rangle-\mu(\langle\psi^{\dagger}\psi\rangle-n_{0}).
\end{align}

A general spin diagonal quartic interaction term in $\langle 
H_{\mathrm{int}}\rangle$ can be
written as
\begin{align}
&\langle(\psi^{\dagger}O_{a}\psi)(\psi^{\dagger}O_{b}\psi)\rangle\nonumber\\
 &=
\langle\psi^{\dagger}O_{a}\psi\rangle\langle\psi^{\dagger}O_{b}
\psi\rangle-\sum_{s}\langle\psi_{\alpha s}^{\dagger}\psi_{\beta' s}
\rangle\langle\psi_{\beta s}^{\dagger}\psi_{\alpha' s}\rangle
O_{a}^{\alpha\alpha'}O_{b}^{\beta\beta'}
\end{align}

Summation over repeated indices is implied. Let us write
$G_{s}^{\alpha\beta}\equiv\langle\psi_{\alpha s}^{\dagger}\psi_{\beta s}
\rangle=\sum_{a}\phi_{a s}O_{a}^{\alpha\beta}$,
where 
$\phi_{a s}=\frac{1}{4}\mbox{Tr}O_{a}G_{s}=\langle\psi_{s}^{
\dagger}O_{a}\psi_{s}\rangle$.
Then

\begin{align}
&\langle\psi_{\alpha s}^{\dagger}\psi_{\beta' s}\rangle\langle\psi_{
\beta s}^{\dagger}\psi_{\alpha' s}\rangle 
O_{a}^{\alpha\alpha'}O_{b}^{\beta\beta'}\nonumber\\
& =\frac{1}{16}\sum_{c,d}\phi_{
c s}\phi_{d s}O_{a}^{\alpha\alpha'}O_{b}^{\beta\beta'}O_{
c}^{\alpha\beta'}O_{d}^{\beta\alpha'}\nonumber\\
& =\frac{1}{16}\sum_{c,d
}\phi_{c s}\phi_{d s}\text{Tr}[O_{a}O_{c}^{*}O_{b}O_{d}^{*}]
\end{align}
For the general local $H_{int}$ we will have $a=b$ which gives 
$c=d$.
Then $\text{Tr}[O_{a}O_{c}^{*}O_{a}O_{c}^{*}]=\pm4$ and
\begin{align}
\langle(\psi^{\dagger}O_{a}\psi)^{2}\rangle= &
\langle\psi^{\dagger}O_{a}\psi\rangle^{2}-\frac{1}{4}\sum_{c,s}\phi_{
c,s } ^{2}\text{Sgn}[O_{a}O_{c}]
\end{align}
Here $\text{Sgn}[O_{a}O_{c}]=+1$ if $O_{a},\; O_{c}$ commute
and $-1$ if they anti-commute. We can then collect and write all the terms 
in a compact way as 
\begin{align}
 \langle H_{\mathrm{int}}\rangle &
=-\frac{1}{2}\sum_{a}\phi_{a}^{T}M_{a}\phi_{a}
\end{align}
where $M_a$'s are $2 \times 2$ matrices in spin label for each $\phi_a$. The 
energy 
functional then becomes

\begin{align}
E(\{\lambda_{a}\},\mu_{0s}) & =\langle 
H_{MF}\rangle+\sum_{a=1}^{15}\lambda_{a}\langle\psi^{\dagger}O_{a}
\psi\rangle+\mu_{0s}
\langle\psi^{\dagger}_s\psi_s\rangle\nonumber \\
 & \hspace{2mm} -\frac{1}{2}\sum_{a=0}^{15}\phi_{a}^{T}M_{
a}\phi_{a}-\mu(\langle\psi^{\dagger}\psi\rangle-n_{0}
)\nonumber \\
 & =\langle 
H_{MF}\rangle+\sum_{a=1}^{15}\lambda_{a}\phi_{a}+\mu_{0s}\langle\psi_s^{\dagger}
\psi_s\rangle \nonumber \\
 & \hspace{2mm}-\frac{1}{2}\sum_{a=0}^{15}\phi_{a}^{T}M_{a}\phi_{a}
-\mu(\langle\psi^{
\dagger}\psi\rangle-n_{0})\label{eq:E}
\end{align}

The saddle point equations are

\begin{align}
\frac{\partial 
 E(\{\lambda_{a}\},\mu_{0s})}{\partial\lambda_{a's'}}& 
=\sum_{a=1}^{15}\left(\lambda_{a}^T-\phi_{
a}^T M_{a}\right)\frac{\partial\phi_{a}}{\partial\lambda_{a's'}
}-\phi_{0}^T M_{0}\frac{\partial\phi_{0}}{\partial\lambda_{a's'}
}\nonumber \\
 & \hspace{9mm}+(\mu_{0s}-\mu)\frac{
\partial\langle\psi_{s}^{\dagger}\psi_{s}\rangle}{\partial\lambda_
{a's'}}=0
\end{align}

\begin{align}
\frac{\partial 
E(\{\lambda_{a}\},\mu_{0s})}{\partial\mu_{0s'}}&=\sum_{a=1}^{15}\left(\lambda_{
a }
^T -\phi_{a}^T 
M_{a}\right)\frac{\partial\phi_{a}}{\partial\mu_{0s'}}-\phi_{
0}^T M_{0}\frac{\partial\phi_{0}}{\partial\mu_{0s'}
}\nonumber \\
 & \hspace{9mm}+(\mu_{0s}-\mu)\frac{\partial\langle\psi_
{s}^{\dagger}\psi_{s} 
\rangle}{\partial\mu_{0s'}}=0
\end{align}

\begin{equation}
\frac{\partial 
E(\{\lambda_{a}\},\mu_{0})}{\partial\mu}=\langle\psi^{\dagger}\psi\rangle-n_{0}
=0
\end{equation}

This gives the following mean-field equations:

\begin{align}
\lambda_{a} & =M_{a}\phi_{a}\nonumber \\
n_{0} & =\langle\psi^{\dagger}\psi\rangle\\
\mu_{0s} &= \mu +M_0 \phi_0
\end{align}

Substituting $\phi_{a}=M_{a}^{-1}\lambda_{a}$ and
$\langle\psi^{\dagger}\psi\rangle=n_{0}$ back into the expression
for the energy, we get

\begin{align}
\tilde{E}(\{\lambda_{a}\},\mu_{0})=&\langle 
H_{MF}\rangle+\frac{1}{2}\lambda_{a}^{T}M_{a}^{-1}
\lambda_{a}+\mu n_0\nonumber \\
 & +\frac{1}{2}(\mu_{0s}-\mu) (M_0^{-1})_{s,s'}(\mu_{0s'}-\mu) 
\label{eq:Aniso}
\end{align}

Note that this energy functional coincides with the original one at
the saddle point, and its variation with respect to $\lambda_{a}$
and $\mu_{0}$ gives the correct mean-field equations.

\subsection{Symmetries of the BLG Hamiltonian}\label{apdx:A}
The non-interacting Hamiltonian of BLG in absence of external magnetic field 
has the following symmetries. Below we describe the behaviour of our 
wavefunctions 
$|\psi_{\tau_z = K/K', \mu_z = \pm 1, \sigma_z = A/B}(\vec{q})\rangle$ under 
the symmetry operations. ($\tau$, $\mu$, $\sigma$ represent Pauli matrices 
acting in valley, layer, and sub lattice space, respectively, and $\vec{K}= 
\left(-\frac{4\pi}{3 \sqrt{3} a_0},0\right)$) where $a_0$ is the 
inter-atomic spacing in each layer. Other definitions as shown in the figure 
are $ \vec{R}_1= 
\left(0,-a_0 \right),\,\vec{R}_2= a_0 \left( - 
\frac{\sqrt{3}}{2},\frac{1}{2}\right),\, \vec{R}_3= a_0 \left(
\frac{\sqrt{3}}{2},\frac{1}{2}\right);\;\; \vec{G}_1= \frac{2 \pi}{a_0} 
\left( \frac{-1}{\sqrt{3}},\frac{1}{3}\right),\, \vec{G}_2= \frac{2 \pi}{a_0} 
\left( \frac{1}{\sqrt{3}},\frac{1}{3}\right).$
\\
\\
1. Time reversal symmetry (TRS): $(\vec{q}\rightarrow -\vec{q}) i s_y \tau_x 
\mathcal{K},\;$   ($\mathcal{K}$ represents complex conjugation). \\
\\
2. Rotation by $2 \pi/3$ around the A-sublattice of top layer (the stacking 
point):
\begin{align*}
&J_{2\pi/3}\exp\!\left[\!\frac{2\pi i}{3} \tau_z \!
\left(\frac{1+\mu_z}{2} \frac{1+\sigma_z}{2}\right)\!\!-\!\frac{2\pi 
i}{3}\tau_z 
\!\!\left(\frac{1-\mu_z}{2} \frac{1-\sigma_z}{2}\right)\! \!\right]\\
& = J_{2\pi/3} 
\exp[\frac{2\pi i}{3}\tau_z 
\left(\frac{\sigma_z+\mu_z}{2}\right)].
\end{align*}
(Under $J_{2\pi/3}, \: 
q_x\rightarrow q_x \cos(2\pi/3)-q_y \sin(2\pi/3),\, q_y\rightarrow q_x 
\sin(2\pi/3)+q_y \cos(2\pi/3)$)\\
\\
3. Translation by $\vec{R}= l_1 \vec{n}_1 + l_2 \vec{n}_2: \; 
\exp[i\tau_z \frac{4\pi}{3 \sqrt{3} a_0}R_x] \exp[-i \vec{q}.\vec{R}],$. 
$\vec{n}_1$ and $\vec{n}_2$ are primitive vectors of the hexagonal lattice 
(see Fig. 2).\\
\\
4. Inversion Symmetry:  $(\vec{q}\rightarrow -\vec{q}) \tau_x \mu_x \sigma_x$, 
with 
inversion center at mid point of the stacking bond between the two layers.\\
\\
5. Mirror $x\rightarrow -x$: $(q_x\rightarrow -q_x)  \tau_x.$
\\
\\
\begin{figure}
\includegraphics[width=7cm]{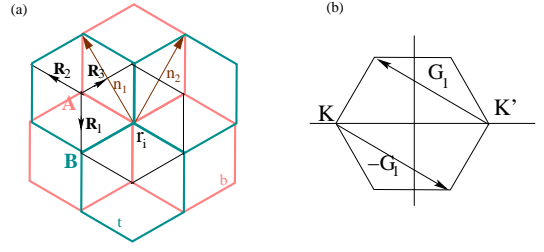}\label{symm1}
\caption{(a) Schematic of bilayer graphene with top/bottom layer labelled by 
t/b (with colours blue/pink). Thin black hexagon defines the effective 
hexagonal lattice. (b) Brilluion zone for the effective hopping Hamiltonian}
\end{figure}
\textbf{Details of the symmetry transformations:}

Rotation by $2\pi/3$ for $|\psi_{K, A}(\vec{q})\rangle$:
% \vspace{-7mm}
\begin{align*}
 \langle \vec{r}_i|\psi_{K, A}(\vec{q})\rangle & \sim \exp[i 
(\vec{K}+\vec{q}).\vec{r}_{i,A}] \\
 \langle \vec{r}_i|J_{\frac{2\pi}{3}}|\psi_{K, A}(\vec{q})\rangle & \sim 
\exp[i 
J_{\frac{2\pi}{3}} (\vec{K}+\vec{q}).\vec{r}_{i,A}]\\
&= \exp[i (\vec{K}-\vec{G}_1 + 
J_{\frac{2\pi}{3}} \vec{q}).(\vec{r}_i+\vec{R}_2)]\\
& =\exp[- i \vec{G}_1.\vec{R}_2]  \langle \vec{r}_i|\psi_{K, 
A}(J_{\frac{2\pi}{3}} \vec{q})\rangle\\
&= \exp\left(i 2\pi/3 \right) \langle 
\vec{r}_i|\psi_{K, 
A}(J_{\frac{2\pi}{3}} \vec{q})\rangle
\end{align*}

Similarly at B sublattice,
\begin{align*}
 \langle \vec{r}_i|J_{\frac{2\pi}{3}}|\psi_{K,B}(\vec{q})\rangle & \sim \exp[i 
J_{\frac{2\pi}{3}} (\vec{K}+\vec{q}).\vec{r}_{i,B}] \\
&= \exp[i 
\vec{K}-\vec{G}_1 
+ J_{\frac{2\pi}{3}} \vec{q}).(\vec{r}_i-\vec{R}_3)]\\
& =\exp[i \vec{G}_1.\vec{R}_3]  \langle \vec{r}_i|\psi_{K, 
A}(J_{\frac{2\pi}{3}} \vec{q})\rangle \\
&= \exp\left(-i2\pi/3 \right)  \langle 
\vec{r}_i|\psi_{K, 
A}(J_{\frac{2\pi}{3}} \vec{q})\rangle.
\end{align*}

At $K',$ we get $\vec{K}'\rightarrow \vec{K}'+\vec{G}_1$ and the signs are 
reversed for $A$ and 
$B$ sublattices. Thus we get the above expression for the symmetry operation.\\
\\
Translation by a lattice translation $\vec{R}$:
% \vspace{-7mm}
\begin{align*}
\langle \vec{r}_i|T_{\vec{R}} \,\psi_{K, A/B}(\vec{q})\rangle & = \langle 
T_{-\vec{R}} \vec{r}_i| 
\,\psi_{K, A/B}(\vec{q})\rangle\\
& \!\!\!\!\!\!\!\!\!\!\!\!\!\!\!\!\!\!\!\!\sim \exp[i ( 
\vec{K}+\vec{q}).(\vec{r}_{i,A/B}-\vec{R})]\\
&\!\!\!\!\!\!\!\!\!\!\!\!\!\!\!\!=\exp[-i (\vec{K}+\vec{q}).\vec{R}]  \langle 
\vec{r}_i|\psi_{K, 
A/B}(\vec{q})\rangle \\
&\!\!\!\!\!\!\!\!\!\!\!\!\!\!\!\!=\exp[i \frac{4\pi}{3 \sqrt{3}a_0}R_x] \exp[-i 
\vec{q}.\vec{R}] \langle 
\vec{r}_i|\psi_{K, 
A/B}(\vec{q})\rangle.
\end{align*}

Similarly at $\vec{K}'(=-\vec{K}); \;\;  \langle \vec{r}_i|T_R \,\psi_{K', 
A/B}(\vec{q})\rangle = 
\exp[-i \frac{4\pi}{3 \sqrt{3}a_0}R_x] \exp[-i \vec{q}.\vec{R}] \langle 
\vec{r}_i|\psi_{K',A/B}(\vec{q})\rangle$.

\subsection{Most general interaction Hamiltonian}

For BLG, we can write all possible symmetry allowed local 
interactions making an analysis similar to that of Vafek 
\textit{et. al.}~\cite{Vafek2010a}.
There are $9$ symmetry allowed spin diagonal terms. We write them in following 
particular form. 

\begin{align}
I_{1}= & (\psi^{\dagger}\psi)^{2}\nonumber \\
I_{2}= & (\psi^{\dagger}\mu^{z}\tau^{z}\psi)^{2}\nonumber \\
I_{3}= &
(\psi^{\dagger}\mu^{x}\psi)^{2}+(\psi^{\dagger}\mu^{y}\tau^{z}\psi)^{2}
\nonumber 
\\
I_{4}= &
(\psi^{\dagger}\mu^{z}\psi)^{2}+(\psi^{\dagger}\mu^{y}\tau^{y}\psi)^{2}+(\psi^{
\dagger
}\mu^{y}\tau^{x}\psi)^{2}\nonumber \\
I_{5}= &
(\psi^{\dagger}\tau^{z}\psi)^{2}+(\psi^{\dagger}\mu^{x}\tau^{x}\psi)^{2}+(\psi^{
\dagger}\mu^{x}\tau^{y}\psi)^{2}\nonumber \\
I_{6}= &
(\psi^{\dagger}\mu^{x}\tau^{z}\psi)^{2}+(\psi^{\dagger}\mu^{y}\psi)^{2}+(\psi^{
\dagger
}\tau^{x}\psi)^{2}+(\psi^{\dagger}\tau^{y}\psi)^{2}\nonumber \\
 & \hspace{1mm} +(\psi^{\dagger}\mu^{z}\tau^{
x}
\psi)^{2}+(\psi^{\dagger}\mu^{z}\tau^{y}\psi)^{2}\nonumber \\
I_{7}= &
(\psi^{\dagger}\mu^{z}\psi)^{2}-(\psi^{\dagger}\mu^{y}\tau^{y}\psi)^{2}-(\psi^{
\dagger
}\mu^{y}\tau^{x}\psi)^{2}\nonumber \\
I_{8}= &
(\psi^{\dagger}\tau^{z}\psi)^{2}-(\psi^{\dagger}\mu^{x}\tau^{x}\psi)^{2}-(\psi^{
\dagger}\mu^{x}\tau^{y}\psi)^{2}\nonumber \\
I_{9}= &
(\psi^{\dagger}\mu^{x}\tau^{z}\psi)^{2}+(\psi^{\dagger}\mu^{y}\psi)^{2}-(\psi^{
\dagger
}\tau^{x}\psi)^{2}-(\psi^{\dagger}\tau^{y}\psi)^{2}\nonumber \\
 & \hspace{1mm}-(\psi^{\dagger}\mu^{z}\tau^{
x}
\psi)^{2}-(\psi^{\dagger}\mu^{z}\tau^{y}\psi)^{2}\\
\nonumber \\
\nonumber 
\end{align}

These combinations of different terms are made such that the 
first six interaction terms above commute with the unitary the transformation 
$\hat{U}=\exp(i\frac{\pi}{4}\mu^{x}\tau^{y}),$ which maps FLP state order 
parameter to STF state. Thus, having only first 6 terms will not lift the 
degeneracy between FLP and STF ground states. 

In addition, for a spin symmetric interaction, the allowed spin dependent terms 
are
$I_{1s}=(\psi^{\dagger}\vec{s}\psi)^{2}$,$I_{2s}=(\psi^{\dagger}\mu^{z}\tau^{z}
\vec{s}\psi)^{2},...I_{9s}.$ 

We can use Fierz identities to find the number of independent interaction 
terms out of the above 18 terms. These identities can be written in terms of 
the 32 individual quartic terms appearing in $I_1,...I_9$ and 
$I_{1s},...I_{9s}$ as: 
\begin{align*}
(\psi^\dagger O_{a}\psi)^{2}= &
-\frac{1}{8}\,\sum_{b}\text{{Sgn}}[O_{a}O_{b}]\,((\psi^\dagger 
O_{b}\psi)^{2}+(\psi^\dagger O_{b}\vec{s}\psi)^{2})\\
(\psi^\dagger O_{a}\vec{s}\psi)^{2}= & 
-\frac{1}{8}\,\sum_{b}\text{{Sgn}}[O_{a}O_{b}]\,(3(\psi^\dagger 
O_{b}\psi)^{2}-(\psi^\dagger O_{b}\vec{s}\psi)^{2})
\end{align*}
\begin{align*}
O_a =&\{1, \mu^z \tau^z,\mu^{x},\mu^{y}\tau^{z}, 
 \,\mu^{z},\mu^{y}\tau^{y},\mu^{y}\tau^{x},\tau^{z} 
,\mu^{x}\tau^{x},\mu^{x}\tau^{y},\nonumber \\
 & \hspace{3mm}\mu^{x}\tau^{z},\mu^{y},\tau^{x},\tau^{y},\mu^{
z}\tau^{x},\mu^ {z}\tau^{y} \}
\end{align*}

Out of these 18 equations, only 9 are independent and therefore we can express 
$(\psi^\dagger O_{a}\vec{s}\psi)^{2}$ terms in terms of $(\psi^\dagger 
O_{a}\psi)^{2}$. Then $I_{1s},
I_{2s}..I_{9s}$ can be rewritten in terms of $I_{1},I_{2}..I_{9}$ 
 as $I_{is}=\sum\,\Gamma_{ij}I_{j}$,
where 

\begin{align}
\Gamma & =\left(\begin{array}{ccccccccc}
-\frac{3}{2} & -\frac{1}{2} & -\frac{1}{2} & -\frac{1}{2} & -\frac{1}{2} &
-\frac{1}{2} & 0 & 0 & 0\\
-\frac{1}{2} & -\frac{3}{2} & \frac{1}{2} & -\frac{1}{2} & -\frac{1}{2} & 
\frac{1}{2}
& 0 & 0 & 0\\
-1 & 1 & -1 & 1 & -1 & 0 & 0 & 0 & 0\\
-\frac{3}{2} & -\frac{3}{2} & \frac{3}{2} & -\frac{1}{2} & \frac{1}{2} & 
-\frac{1}{2}
& 0 & 0 & 0\\
-\frac{3}{2} & -\frac{3}{2} & -\frac{3}{2} & \frac{1}{2} & -\frac{1}{2} & 
\frac{1}{2}
& 0 & 0 & 0\\
-3 & 3 & 0 & -1 & 1 & -1 & 0 & 0 & 0\\
\frac{1}{2} & \frac{1}{2} & -\frac{1}{2} & -\frac{1}{2} & -\frac{1}{2} & 
\frac{1}{2} &
-2 & -1 & 1\\
\frac{1}{2} & \frac{1}{2} & \frac{1}{2} & -\frac{1}{2} & -\frac{1}{2} & 
-\frac{1}{2} &
-1 & -2 & -1\\
1 & -1 & 0 & 1 & -1 & 0 & 2 & -2 & -1
\end{array}\right). \label{eq:spin_fierz}
\end{align}

The most general local interaction Hamiltonian for BLG can now be written as
\begin{align}
H_{\text{int}} & =\sum_{i=1}^{9}g_{i}I_{i} \label{eq:gen_int}
\end{align}

\section{Mean field analysis for $\nu=2$ BLG}\label{apdx:D}
For filling fraction $\nu=2$, we consider two symmetry broken states and compare
their ground state energies to find the favored state at zero electric field. 
Below
we discuss the the mean field analysis for both of them.

\subsection{Layer polarized state}

We first consider a state in which inversion symmetry is broken, but
translational and rotational symmetries are preserved. In such state,
the following mean-field terms are allowed:

\begin{equation}
O_{a=\{1,2,3\}}=\{\mu^{z},\tau^{z},\mu^{z}\tau^{z}\}.
\end{equation}

Note that, since we have assumed a uniform state and these are local
terms, they must be diagonal in the LL index $n$. We therefore write
the mean-field Hamiltonian as follows:

\begin{equation}
H_{MF}=H_{0}-\!\!\!\!\!\!\!\sum_{\zeta=\pm1,a=1,2,3}\!\!\!\!\!\lambda_{a\zeta}
\psi^ { \dagger}
\frac{1+\zeta 
s^{z}}{2}O_{a}\psi-h\psi^{\dagger}s^{z}\psi-\mu_{0}\psi^{\dagger}\psi,\label{
eq:HMF-1}
\end{equation}

where $h$ is a Zeeman field. The Hamiltonian can be diagonalized and the 
spectrum is

\begin{align}
E_{n,s,\tau} &\!\!=\begin{cases}
-hs-\mu_{0}-(\lambda_{1s}+\lambda_{2s})\tau-\lambda_{3s}\!\!\!\!\! & ,\mbox{ 
}n\le1\\
-hs-\mu_{0}-\lambda_{2s}\tau\pm\sqrt{(\lambda_{1s}+\tau\lambda_{3s})^{2}+E_{n}^{
2}}\!\!\!\!\! & ,\mbox{ }n>1
\end{cases}
\end{align}

Filling all the Landau levels up to $\nu=2$, the total energy is:

\begin{align}
&E_{MF} 
=\sum_{\tau=\pm1}\left(E_{0,\uparrow,\tau}+E_{1,\uparrow,\tau}\right)+E_{0,
\downarrow,\tau=1}+E_{1,\downarrow,\tau=1}\nonumber \\
 & \hspace{12mm} +\sum_{n=2}^{\infty}\sum_{\tau,s=\pm1}
E_{n,s
,\tau} \nonumber\\
 &
=2\left[-h-2\lambda_{3\uparrow}-\lambda_{1\downarrow}-\lambda_{2\downarrow}
-\lambda_{
3\downarrow}\right]\nonumber \\
 & \hspace{3mm}-\sum_{n=2}^{\infty}\sum_{\tau,s=\pm1}\sqrt{(\lambda_{1s}
+\tau\lambda_{3s})^{2}+E_{n}^{2}} \label{eq:flp_emf}
\end{align}

We first consider the weak interaction limit such that $\lambda \ll \hbar 
\omega_c$
and we can expand to second order in the $\lambda$'s. We get (ignoring the 
chemical
potential terms)

\begin{align}
E_{MF} 
 & \approx
2\left[-h-2\lambda_{3\uparrow}-\lambda_{1\downarrow}-\lambda_{2\downarrow}
-\lambda_{
3\downarrow}\right]\nonumber \\
 & \hspace{5mm}-\frac{\chi_{0}\ell_{B}^2}{2}\sum_{s=\pm1}\left(\lambda_{1s}^
{2}+\lambda_{3s}
^{2}\right)-E_{0}
\end{align}
where 
$\chi_{0}\equiv\sum_{n=2}^{\infty}\frac{2}{E_{n}\ell_{B}^2}=\frac{m}{\pi}\sum_{
n=2 } ^ { \infty } \frac{1}{\sqrt{n(n-1)}}$ and 
$E_{0}=-4\sum_{n=2}^{\infty}E_{n}$ 
(Note that these sums actually diverge, and a cutoff needs to be introduced). 
$E_{MF}$ above is the energy per $\ell_{B}^2$ area of the system while 
$\langle H_{int} \rangle$ terms are energy per unit area. 
Thus, collecting the different terms in Eq. (\ref{eq:Aniso}) and matching the 
dimensions, we get (dropping constants)

\begin{align}
& \tilde{E}(\lambda_{\gamma},\mu_{0})  
=\frac{1}{\ell_B^2}(-4\lambda_{3\uparrow}-2\lambda_{1\downarrow}-2\lambda_{
2\downarrow } -2\lambda_ { 3\downarrow})\nonumber \\
 &
+\frac{1}{2}\lambda_{1}^{T}\left(M_{1}^{-1}-\chi_0
\right)\lambda_{1}+\frac{1}{2}
\lambda_{2}^{T}M_{2}^{-1}\lambda_{2}+\frac{1}{2}\lambda_{3}^{T}\left(M_{3}^{-1}
-\chi_0\right)\lambda_{3} \label{eq:e_flp}
\end{align}

This is conveniently written as

\begin{equation}
\tilde{E}(\lambda_{\gamma},\mu_{0})=-\frac{1}{\ell_B^2}Q^{T}\lambda+\frac{1}{2}
\lambda^{T}\tilde{M}^{-1}\lambda,\label{eq:Emin}
\end{equation}

where 
$\lambda^{T}=(\lambda_{1\uparrow},\lambda_{1\downarrow},\dots,\lambda_{
3\downarrow})$.
Minimizing the energy over $\lambda$ gives

\begin{align}
E_{min}&=-\frac{1}{2 \ell_B^4}Q^{T}\tilde{M}Q \label{eq:emin_flp}
\end{align}

We can now calculate $\langle H_{int}\rangle_{FLP},$ using the general 
interaction
Hamiltonian in Eq.~(\ref{eq:gen_int}). In FLP, three $\phi$'s corresponding to
the above $O_{a}$'s are nonzero. Calculating $\langle H_{int}\rangle_{FLP},$ we
get the following $M_i$ matrices in Eq.~(\ref{eq:e_flp}) for FLP phase: 

\begin{equation}
M_{1}=\left(\begin{array}{cc}
\frac{K_{1}}{2}-2g_{4}-2g_{7} & -2g_{4}-2g_{7}\\
\\
-2g_{4}-2g_{7} & \frac{K_{1}}{2}-2g_{4}-2g_{7}
\end{array}\right),
\end{equation}

\begin{equation}
M_{2}=\left(\begin{array}{cc}
\frac{K_{2}}{2}-2g_{5}-2g_{8} & -2g_{5}-2g_{8}\\
\\
-2g_{5}-2g_{8} & \frac{K_{2}}{2}-2g_{5}-2g_{8}
\end{array}\right),
\end{equation}
\begin{equation}
M_{3}=\left(\begin{array}{cc}
\frac{K_{3}}{2}-2g_{2} & -2g_{2}\\
\\
-2g_{2} & \frac{K_{3}}{2}-2g_{2}
\end{array}\right),
\end{equation}

where 

$K_{1}=g_{1}+g_{2}-2g_{3}-g_{4}-g_{5}+2g_{6}+3g_{7}+3g_{8}-6g_{9}$

$K_{2}=g_{1}+g_{2}+2g_{3}-g_{4}-g_{5}-2g_{6}+3g_{7}+3g_{8}+6g_{9}$

$K_{3}=g_{1}+g_{2}-2g_{3}+3g_{4}+3g_{5}-6g_{6}-g_{7}-g_{8}+2g_{9}$

\subsection{Staggered flux state}

The second natural possibility for an ordered state (that lifts the
degeneracy of the lowest Landau level) is a staggered flux (STF) state
with a wavevector that connects $K$ to $K'$. This state preserves
the three fold rotational symmetry $\hat{R}_{2\pi/3}=\exp(2\pi 
i\mu^{z}\tau^{z}/3)$ around
the aligned sites and the inversion symmetry, but breaks translational
symmetry. The allowed mean fields in this state are

\begin{equation}
O_{a=\{1,2,3\}}=\{\mu^{x}\tau^{x},\mu^{y}\tau^{y},\mu^{z}\tau^{z}\}.
\end{equation}

Now,
\begin{align}
H_{MF}=H_{0}-\!\!\!\!\!\!\!\!\!\!\sum_{\zeta=\pm1,a=1,2,3}\!\!\!\lambda_{a\zeta}
\psi^{ \dagger} \frac {
1+\zeta 
s^{z}}{2}O_{a}\psi-h\psi^{\dagger}s^{z}\psi-\mu_{0}\psi^{\dagger}\psi
\end{align}
which has the spectrum
\begin{align}
E_{n,s,\tau} &\!\! =\begin{cases}
-hs-\mu_{0}+(\lambda_{1s}-\lambda_{2s})\tau-\lambda_{3s}\!\!\!\!\!\! & ,\mbox{ 
}n\le1\\
-hs-\mu_{0}+\lambda_{1s}\tau\pm\sqrt{(\lambda_{2s}+\tau\lambda_{3s})^{2}+E_{n}^{
2}}\!\!\!\!\!\! &
,\mbox{ }n>1.
\end{cases}
\end{align}
Filling all the Landau levels up to $\nu=2$, the total energy is:

\begin{align}
E_{MF}&=\sum_{\tau=\pm1}\left(E_{0,\uparrow,\tau}+E_{1,\uparrow,\tau}\right)+E_{
0,
\downarrow,\tau=1}+E_{1,\downarrow,\tau=1}\nonumber \\
 & \hspace{6mm}+\sum_{n=2}^{\infty}\sum_{\tau,s=\pm1}
E_{n,s
,\tau}\nonumber\\
&
=2\left[-h-|\lambda_{1\downarrow}-\lambda_{2\downarrow}|-2\lambda_{3\uparrow}
-\lambda_
{3\downarrow}\right]\nonumber \\
 & \hspace{3mm}-\sum_{n=2}^{\infty}\sum_{\tau,s=\pm1}\sqrt{(\lambda_{2s}
+\tau\lambda_{3s})^{2}+E_{n}^{2}} \label{eq:emf_stf}
\end{align}

Again considering the weak interaction limit and expanding to second order in 
the
$\lambda$'s, we get

\begin{align}
E_{MF}  &
\approx 2\left[-h-|\lambda_{1\downarrow}-\lambda_{2\downarrow}
|-2\lambda_{3\uparrow}
-\lambda_{3\downarrow}\right]\nonumber \\
 & \hspace{6mm}-\frac{\chi_{0}\ell_B^2}{2}\sum_{s=\pm1}
\left(\lambda_ { 2s } ^{2}
+\lambda_{3s}^{2}\right)-E_{0}
\end{align}

Taking $\langle H_{int}\rangle_{STF},$ we get the following $M_i$ matrices
corresponding to the STF order parameters above

\begin{equation}
M_{1S}=\left(\begin{array}{cc}
\frac{K_{1S}}{2}-2g_{5}+2g_{8} & -2g_{5}+2g_{8}\\
\\
-2g_{5}+2g_{8} & \frac{K_{1S}}{2}-2g_{5}+2g_{8}
\end{array}\right),
\end{equation}

\begin{equation}
M_{2S}=\left(\begin{array}{cc}
\frac{K_{2S}}{2}-2g_{4}+2g_{7} & -2g_{4}+2g_{7}\\
\\
-2g_{4}+2g_{7} & \frac{K_{2S}}{2}-2g_{4}+2g_{7}
\end{array}\right),
\end{equation}
\begin{equation}
M_{3S}=\left(\begin{array}{cc}
\frac{K_{3S}}{2}-2g_{2} & -2g_{2}\\
\\
-2g_{2} & \frac{K_{3S}}{2}-2g_{2}
\end{array}\right),
\end{equation}

where 

\begin{align*}
K_{1S}= & g_{1}+g_{2}+2g_{3}-g_{4}-g_{5}-2g_{6}-g_{7}-g_{8}-2g_{9}\\
K_{2S}= & g_{1}+g_{2}-2g_{3}-g_{4}-g_{5}+2g_{6}-g_{7}-g_{8}+2g_{9}\\
K_{3S}= & g_{1}+g_{2}-2g_{3}+3g_{4}+3g_{5}-6g_{6}-g_{7}-g_{8}+2g_{9}
\end{align*}

\begin{align}
\tilde{E}(\lambda_{\gamma},\mu_{0}) &
=\frac{2}{\ell_B^2}\left[-h-(\lambda_{1\downarrow}-\lambda_{2\downarrow}
)-2\lambda_{3\uparrow}
-\lambda_{3\downarrow}\right]+\frac{1}{2}\lambda_{1}^{T} 
M_{1S}^{-1}\lambda_{1}\nonumber \\
 & \hspace{2mm}+\frac{1}{2}
\lambda_{2}^{T}\left(M_{2S}^{-1}-\chi_0\right)\lambda_{
2 } +\frac{1}{2}\lambda_{3}^{T}
\left(M_ { 3S} ^ { -1}-\chi_0\right)\lambda_{3} \nonumber \\
&=-\frac{1}{\ell_B^2}Q_S^{T}\lambda+\frac{1}{2}\lambda^{T}\tilde{M_S}^{-1
}
\lambda,\label{eq:e_stf}
\end{align}

And after minimizing over $\lambda$

\begin{equation}
E_{min}=-\frac{1}{2\ell_B^4}Q_S^{T}\tilde{M_S}Q_S. \label{eq:emin_stf}
\end{equation}

Using equations (\ref{eq:emin_flp}) and (\ref{eq:emin_stf}), we can calculate 
the
ground state energies of the two phases ($E_{\text{FLP}}$ and $E_{\text{STF}}$) 
and
compare
which phase has lower energy. We get

\begin{align}
&\Delta E=E_{\text{FLP}}-E_{\text{STF}}  \nonumber \\
 &
=\frac{4 
\chi_{0}}{\ell_B^4}\left(g_{1}g_{7}+g_{2}g_{7}-2g_{3}g_{7}-13g_{4}g_{7 
}-g_{5}g_{7}+2g_{6}g_{7}+g_{7}^{2}\right.\nonumber \\
 & \left.-g_{1}g_{8}-g_{2}g_{8}+2g_{3}g_{8}+5g_{4}g_{8}
+g_{5}g_{8}-2g_{6}g_{8}-g_{8}^{2}+2g_{1}g_{9}\right.\nonumber \\
 & 
\left.+2g_{2}g_{9}-4g_{3}g_{9}-10g_
{4}g_{9}-2g_{5}g_{9}+4g_{6}g_{9}+4g_{8}g_{9}-4g_{9}^{2}\right)\label{eq:deltaE}
\end{align}

\subsection{Large Interactions}

For the case when the interactions are large and above approximation
doesn't work, we numerically minimize the energy functionals 
$E(\{\lambda_{a}\},\mu_{0})$
w.r.t. $\lambda_{a}$. In Eq.~(\ref{eq:E}), we use the MF
equation $\phi_{as}=-\partial E_{MF}/\partial\lambda_{as}$ and find
the minimum of the resulting energy functional.

\section{Change of variables from $V_0, U, V_{1,2,3}$ and $J_H$ to $g_1...g_9$ 
in the Coulomb Hamiltonian}\label{apdx:E}

We have found the ground state energies for the most general interactions in 
the 
weak
coupling limit. We can now use our interaction Hamiltonian in Eq. 
(\ref{eq:hc_int}) and
read off the interaction parameters $g_{1},..g_{9}$. Then we can obtain the 
difference in ground
state energies of FLP and STF phases from Eq. (\ref{eq:deltaE}) to determine 
which 
state is favored.

To write the $U$ term, we use the identity
\begin{equation}
(n_{\uparrow}-\frac{1}{2})(n_{\downarrow}-\frac{1}{2})=\frac{1}{4}-\frac{1}{6}
\left(\psi^{\dagger}\vec{s}\psi\right)^{2}
\end{equation}
that holds for a single orbital.

Up to a chemical potential term this gives,
\begin{align}
&\sum_{\mu}  n_{\mu K\uparrow}n_{\mu K\downarrow}+n_{\mu K'\uparrow}n_{\mu 
K'\downarrow}\nonumber \\
 & 
=-\frac{1}{6}\left[\left(\psi^{\dagger}\frac{1+\mu^{z}}{2}\frac{1+\tau^{z}}{2}
\vec{s}\psi\right)\cdot\left(\psi^{\dagger}\frac{1+\mu^{z}}{2}\frac{1+\tau^{z}}{
2}\vec{s}\psi\right)\right.\nonumber \\
 & \left.\hspace{3mm}+\left(\psi^{\dagger}\frac{1+\mu^{z}}{2}\frac{1-\tau^{z
}}{2}\vec{s}\psi\right)\cdot\left(\psi^{\dagger}\frac{1+\mu^{z}}{2}\frac{
1-\tau^{z}}{2}\vec{s}\psi\right)\right]\nonumber \\
 & \hspace{3mm}+(\mu\rightarrow-\mu)\nonumber \\
\nonumber \\
 & 
=-\frac{1}{6}\left[\frac{1}{4}\left(\psi^{\dagger}\vec{s}
\psi\right)\cdot\left(\psi^{\dagger}\vec{s}\psi\right)+\frac{1}{4}
\left(\psi^{\dagger}\mu^{z}\vec{s}\psi\right)\cdot\left(\psi^{\dagger}\mu^{
z}\vec{s}\psi\right)\right.\nonumber \\
 & \left.+\frac{1}{4}\left(\psi^{\dagger}\tau^{z}\vec{s}
\psi\right)\cdot\left(\psi^{\dagger}\tau^{z}\vec{s}\psi\right)+\frac{1}{4}
\left(\psi^{\dagger}\mu^{z}\tau^{z}\vec{s}\psi\right)\cdot\left(\psi^{
\dagger}\mu^{z}\tau^{z}\vec{s}\psi\right)\right]\nonumber \\
 & 
=-\frac{1}{24}\left(I_{1s}+I_{2s}+\frac{I_{4s}+I_{7s}}{2}+\frac{I_{5s}+I_{8s}}{2
}\right)
\end{align}
Using the the relations between $I_{is}$ and $I_{s}$ in 
Eq. (\ref{eq:spin_fierz}), we
get

\begin{align}
&n_{\mu K\uparrow}n_{\mu K\downarrow}+n_{\mu K'\uparrow}n_{\mu 
K'\downarrow}\nonumber \\
 & 
\hspace{4mm}=\frac{1}{8}\left(I_{1}+I_{2}+\frac{I_{4}+I_{7}}{2}+\frac{I_{5}+I_
{8}}{2}\right).
\end{align}

$V_1$ term:
\begin{align}
&V_1\sum_{\mu}n_{\mu K}n_{\mu K'} \nonumber \\
 &
=V_1\left(\psi^{\dagger}\frac{1+\mu^{z}}{2}\frac{1+\tau^{z}}{2}
\psi\right)\left(\psi^{\dagger}\frac{1+\mu^{z}}{2}\frac{1-\tau^{z}}{2}
\psi\right)+(\mu\rightarrow-\mu) \nonumber \\
 &
=\frac{V_1}{8}\left[\left(\psi^{\dagger}\psi\right) \left(\psi^{\dagger}
\psi\right)+\left(\psi^{\dagger}\mu^{z}\psi\right) \left(\psi^{\dagger}
\mu^{
z}
\psi\right)\right.\nonumber \\
&\hspace{10mm}\left.-\left(\psi^{\dagger}\tau^{z}\psi\right) \left(\psi^
{ \dagger }
\tau^{z}
\psi\right)-\left(\psi^{\dagger}\mu^{z}\tau^{z}\psi\right) \left(\psi^{
\dagger}
\mu^{z}\tau^{z}\psi\right)\right] \nonumber\\
 &
=\frac{V_1}{8}\left(I_{1}-I_{2}+\frac{I_{4}+I_{7}}{2}-\frac{I_{5}+I_{8}}{2}
\right).
\end{align}

$V_{2}$ term:
\begin{align}
 & V_{2}(n_{-1K}n_{1K}+n_{-1K'}n_{1K'})\nonumber \\
 &
=V_{2}\left[\left(\psi^{\dagger}\frac{1+\mu^{z}}{2}\frac{1+\tau^{z}}{2}
\psi\right)\left(\psi^{\dagger}\frac{1-\mu^{z}}{2}\frac{1+\tau^{z}}{2}
\psi\right)\right.\nonumber \\
&\hspace{10mm}\left.+\left(\psi^{\dagger}\frac{1+\mu^{z}}{2}\frac{1-\tau^{z}}{2}
\psi\right)\left(\psi^{\dagger}\frac{1-\mu^{z}}{2}\frac{1-\tau^{z}}{2}
\psi\right)\right]\nonumber \\
 &
=\frac{V_{2}}{8}\left[\left(\psi^{\dagger}\psi\right)\left(\psi^{\dagger}
\psi\right)-\left(\psi^{\dagger}\mu^{z}\psi\right)\left(\psi^{\dagger}\mu^{z}
\psi\right)\right.\nonumber \\
&\hspace{10mm}\left.+\left(\psi^{\dagger}\tau^{z}\psi\right)\left(\psi^{\dagger}
\tau^{z}
\psi\right)-\left(\psi^{\dagger}\mu^{z}\tau^{z}\psi\right)\left(\psi^{
\dagger}\mu^{z}\tau^{z}\psi\right)\right]\nonumber \\
 &
=\frac{V_{2}}{8}\left(I_{1}-I_{2}-\frac{I_{4}+I_{7}}{2}+\frac{I_{5}+I_{8}}{2}
\right).
\end{align}

$V_{3}$ term:
\begin{align}
&V_{3}(n_{-1K}n_{1K'}+n_{-1K'}n_{1K}) \nonumber\\
&=V_{3}\left[\left(\psi^{\dagger}\frac{1+\mu^{z}}{2}\frac{1+\tau^{z}}{2}
\psi\right)\cdot\left(\psi^{\dagger}\frac{1-\mu^{z}}{2}\frac{1-\tau^{z}}{2}
\psi\right)\right.\nonumber \\
&\hspace{10mm}\left.+\left(\psi^{\dagger}\frac{1+\mu^{z}}{2}\frac{1-\tau^{z}}{2}
\psi\right)\cdot\left(\psi^{\dagger}\frac{1-\mu^{z}}{2}\frac{1+\tau^{z}}{2}
\psi\right)\right]\nonumber \\
 &
=\frac{V_{3}}{8}\left[\left(\psi^{\dagger}\psi\right)\left(\psi^{\dagger}
\psi\right)-\left(\psi^{\dagger}\mu^{z}\psi\right)\left(\psi^{\dagger}\mu^{z}
\psi\right)\right.\nonumber \\
&\hspace{10mm}\left.-\left(\psi^{\dagger}\tau^{z}\psi\right)\left(\psi^{\dagger}
\tau^{z}
\psi\right)+\left(\psi^{\dagger}\mu^{z}\tau^{z}\psi\right)\left(\psi^{\dagger}
\mu^{z}
\tau^{z}\psi\right)\right]\nonumber \\
 &
=\frac{V_{3}}{8}\left(I_{1}+I_{2}-\frac{I_{4}+I_{7}}{2}-\frac{I_{5}+I_{8}}{2}
\right).
\end{align}

$J_{H}$ term:
\begin{align}
&-J_{H}\sum_{\mu}\vec{s}_{\mu K}\cdot\vec{s}_{\mu K'} \nonumber \\
&\hspace{2mm} 
=-J_{H}\left[\left(\psi^{\dagger}\frac{1+\mu^{z}}{2}\frac{1+\tau^{z}}{2}\vec{
s}\psi\right)\cdot\left(\psi^{\dagger}\frac{1+\mu^{z}}{2}\frac{1-\tau^{z}}{
2}\vec{s}\psi\right)\right.\nonumber \\
&\hspace{10mm}\left.+(\mu\rightarrow-\mu)\right]\nonumber \\
 & 
=-\frac{J_{H}}{8}\left[\left(\psi^{\dagger}\vec{s}\psi\right)^{2}+\left(\psi^{
\dagger}\mu^{z}\vec{s}\psi\right)^{2}-\left(\psi^{\dagger}\tau^{z}\vec{
s}\psi\right)^{2}-\left(\psi^{\dagger}\mu^{z}\tau^{z}\vec{s}
\psi\right)^{2}\right]\nonumber \\
 & 
=-\frac{J_{H}}{8}\left(I_{1s}-I_{2s}+\frac{I_{4s}+I_{7s}}{2}-\frac{I_{5s}+I_{8s}
}{2}\right)\nonumber \\
 & 
=-\frac{J_{H}}{8}\left(-I_{1}+I_{2}-\frac{I_{4}+I_{7}}{2}+\frac{I_{5}+I_{8}}{2}
-I_{6}+I_{9}\right)
\end{align}

These give the following values of the interaction parameters.

\begin{align}
g_{1} & =V_{0}+\frac{1}{8}\left(U+V_1+V_{2}+V_{3}+J_{H}\right)\nonumber \\
g_{2} & =\frac{1}{8}\left(U-V_1-V_{2}+V_{3}-J_{H}\right)\nonumber \\
g_{3} & =0\nonumber \\
g_{4} & =\frac{1}{16}\left(U+V_1-V_{2}-V_{3}+J_{H}\right)\nonumber \\
g_{5} & =\frac{1}{16}\left(U-V_1+V_{2}-V_{3}-J_{H}\right)\nonumber \\
g_{6} & =\frac{J_{H}}{8}\nonumber \\
g_{7} & =\frac{1}{16}\left(U+V_1-V_{2}-V_{3}+J_{H}\right)\nonumber \\
g_{8} & =\frac{1}{16}\left(U-V_1+V_{2}-V_{3}-J_{H}\right)\nonumber \\
g_{9} & =-\frac{1}{8}J_{H}
\end{align}

Plugging these in Eq. (\ref{eq:deltaE}), we get the difference in mean
field ground state energies of FLP and STF phases for the weak interaction
limit where we can expand the Mean Field energies to $2^{nd}$ order
in $\lambda_{a}$.

\bibliography{BLG-PRB-v3}

\end{document}